\documentclass[runningheads]{llncs}
\usepackage{graphicx}
\usepackage{amsmath}
\usepackage{xcolor}
\usepackage{siunitx}
\usepackage{graphicx}
\usepackage{booktabs}
\usepackage{multirow}
\usepackage{subcaption} % for subfigures
\usepackage{caption}
\begin{document}
\title{Layered Dielectric Characterization of Human Skin in the Sub-Terahertz and Terahertz Frequency Ranges}
\titlerunning{Dielectric Characterization of Human Skin in the Sub-THz and THz Ranges}

\author{Silvia Mura\inst{1}\orcidID{0000-0002-0207-5730} \and Elisabetta Marini\inst{1}\orcidID{0009-0003-1081-7815} \and Maurizio Magarini\inst{1}\orcidID{0000-0001-9288-0452} \and Matti H\"am\"al\"ainen\inst{2}\orcidID{0000-0002-6115-5255}\and
Marco Hernandez\inst{2}}
\authorrunning{S. Mura et al.}
\institute{$^1$ Dept. of Electronics, Information and Bioengineering, Politecnico di Milano, Italy\\
\email{\{elisabetta.marini,silvia.mura,maurizio.magarini\}@polimi.it}\\
$^2$ University of Oulu, Oulu, Finland\\
\email{\{matti.hamalainen,marco.hernandez\}@oulu.fi}}
\maketitle              % typeset the header of the contribution
\begin{abstract}
Sub-terahertz (sub-THz) and terahertz (THz) radiation offer unique opportunities for non-invasive diagnostics and imaging due to their sensitivity to water content and molecular dynamics in biological tissues. In this work, a comprehensive dielectric model of human skin and its cellular constituents is developed across these frequency ranges. The model combines multi-Debye relaxation theory with effective medium formulations to account for intracellular water dynamics and macromolecular relaxation processes. Key cellular parameters, including water content, protein and lipid fractions, and ionic conductivity, are integrated from experimentally validated sources. The proposed framework enables realistic predictions of frequency-dependent permittivity for different skin layers and cell types, providing a physically interpretable description of sub-THz and THz tissue interactions. This approach establishes a foundation for the design and optimization of next-generation diagnostic and imaging techniques operating in these frequency bands.


\end{abstract}

\section{Introduction}
The interaction between sub-terahertz (sub-THz) and terahertz (THz) radiation with biological tissues has gained increased attention for diagnostics, imaging, and sensing applications~\cite{fischer2005,pickwell2006}. In these frequency ranges, electromagnetic waves couple to vibrational and rotational modes, revealing intermolecular forces and collective molecular dynamics. This sensitivity is especially relevant in water-rich biological media, where sub-THz and THz waves enable non-ionizing, non-destructive probing of tissue hydration and cellular water content. While sub-THz radiation penetrates deeper due to lower absorption, THz waves offer higher spectral resolution, making the two ranges complementary in biomedical contexts. 

At the molecular level, weak interactions such as hydrogen bonds and van der Waals forces govern biomolecular structure and functions~\cite{Wang2019NanoBio,Cherkasova2021}. The sub-THz and THz window ($0.1-10\,$THz) coincides with many rotational and vibrational molecular modes, granting access to molecular and collective motions beyond the reach of infrared spectroscopy~\cite{Cherkasova2021}. In aqueous environments, these frequencies are particularly sensitive to hydrogen-bond dynamics, providing insights into the organization and fluctuations of biomolecular systems, because water is the predominant constituent of biological tissue. Consequently, spectroscopy in the sub-THz and THz domains has emerged as a powerful approach to investigate the structural organization and functional dynamics of biomolecular systems~\cite{Lee2025}.

Propagation in biological tissues depends mainly on water absorption, given their 60–80\% water content~\cite{Xu2024,Vilagosh2020}. Scattering also contributes, arising from the structural heterogeneity of tissues, while reflections at interfaces with different refractive indices provide imaging contrast, for instance in distinguishing malignant from healthy tissue~\cite{Xu2024,Vilagosh2020}. Dispersion further modulates wave propagation, influencing time-domain spectroscopy results~\cite{Vilagosh2020}.

Modeling  sub-THz and THz propagation in tissue often rely on frequency-dependent complex permittivity $\varepsilon(\omega) = \varepsilon'(\omega) - j\varepsilon''(\omega)$, with the imaginary part accounting for absorption. Advanced approaches employ numerical simulations or finite element methods in COMSOL to capture realistic geometries, dielectric heterogeneity, and scattering effects~\cite{Mahdy2024,Yanina2022}.

Skin is a key focus for sub-THz and THz studies, being the body’s outermost layer and a target for cancer detection~\cite{Woodward2003,Ferdous2025}, non-invasive imaging~\cite{DArco2020}, and body-centric communications~\cite{ashworth2009}. Its layered structure—epidermis, dermis, and hypodermis—shows distinct dielectric properties linked to hydration, macromolecular content, and lipid composition~\cite{hanson2015}. Therefore, realistic modeling of propagation in these frequency ranges therefore requires a multi-layer description of skin that accounts for its anatomical and cellular heterogeneity. 

The dielectric response of biological cells in these ranges is commonly modeled using Debye-type relaxation. This formalism describes the frequency-dependent complex permittivity as the result of dipolar relaxation processes, which are especially relevant for polar species, such as water, proteins, and lipids~\cite{Pickwell2004}. Multi-Debye formulations account for multiple relaxation processes: the parameters can be tuned according to water content, protein and lipid composition, and ionic conductivity, offering a physically interpretable framework for predicting cellular dielectric properties and constructing realistic tissue-level models. Linking water dynamics and macromolecular contributions to overall permittivity is thus fundamental to advancing sub-THz and THz sensing and imaging in biomedical applications.

In this work, we propose a comprehensive dielectric model of human skin and its cellular constituents across the sub-THz and THz frequency ranges. The framework integrates multi-Debye relaxation theory with effective medium formulations to describe intracellular water and macromolecular inclusions, yielding realistic frequency-dependent permittivity predictions validated against experimental data.

The paper is organized as follows. Section~\ref{sec:skin_model} describes the multilayer skin model and the considered cellular constituents. Section~\ref{sec:cell_permittivity} details the dielectric computation framework. In Sec.~\ref{sec:intrabody_losses}, we analyze intra-body wave propagation. Section~\ref{sec:simulation_framework} describes the voxel-based simulation framework used to model realistic skin tissue scenarios.

\section{Skin Model and Cells}
\label{sec:skin_model}
Human skin is a complex, multilayered organ whose structural and compositional heterogeneity strongly influences its electromagnetic behavior in the millimeter-wave, sub-THz, and THz frequency ranges~\cite{Permittivity2024,SubTHz2024,Hydration2024}. It is composed of multiple layers with distinct water content, protein and lipid composition, and cellular organization, leading to depth-dependent variations in dielectric properties. Understanding these variations is crucial for accurate modeling of sub-THz and THz propagation and for reliable interpretation of imaging and sensing data.

\begin{figure} [b]
    \centering
    \includegraphics[width=0.5\linewidth]{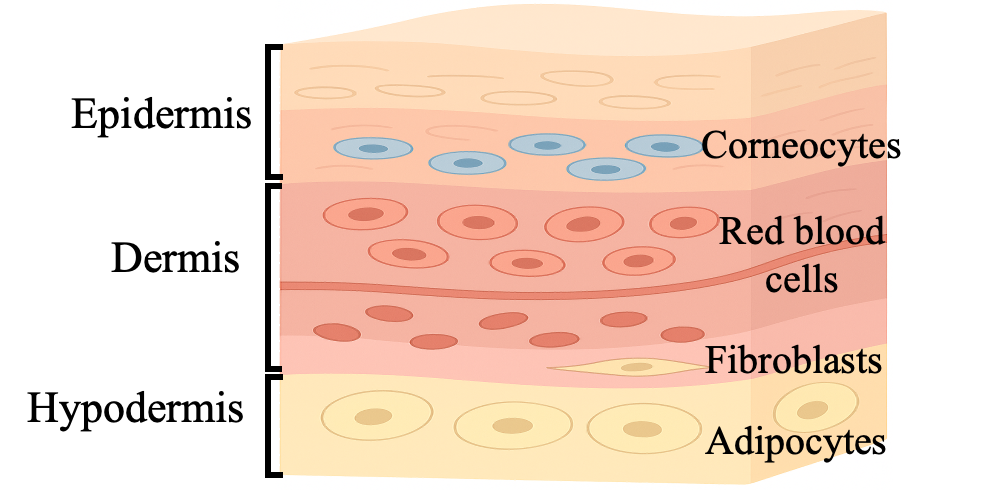}
    \caption{Schematic representation of skin layers and some cell types.}
    \label{skin}
\end{figure}

The overall dielectric response of skin arises from the combined contributions of cellular and extracellular components, including water, proteins, lipids, and ionic species~\cite{pickwell2006,gabriel1996}. In this context, only cellular structures whose dimensions are comparable to the wavelengths or penetration depths of sub-THz and THz radiation are expected to significantly interact with the electromagnetic field, contributing to scattering and absorption phenomena. Modeling the skin as a multilayer system, as illustrated in Fig.~\ref{skin}, enables the representation of these heterogeneous effects on the macroscopic dielectric response. This framework provides a physically grounded basis for simulating sub-THz and THz tissue interactions and for predicting layer-specific electromagnetic contrasts.

\subsection{Epidermis}
The epidermis is the outermost skin layer and the primary interface with the external environment. Its thickness can reach up to $600\,\mu\text{m}$, and comprises five tightly interconnected strata. Lacking blood vessels, it depends on diffusion from the underlying dermis for nutrient and waste exchange. Functionally, it provides mechanical protection, prevents water loss, and supports immune defense~\cite{fluhr2024,tang2024,Hofmann2023}. The epidermis is primarily composed of specialized cell types~\cite{cells2023}. The \textbf{keratinocytes}, the predominant cell population (90--95\%), form the structural scaffold and produce keratin. Originating in the basal layer, they migrate outward through progressive differentiation stages that maintain tissue homeostasis: the \textit{basal keratinocytes} in the stratum basale, are proliferative cells responsible for continuous renewal; the \textit{spinous keratinocytes} in the stratum spinosum, provide mechanical strength; and the \textit{granular keratinocytes} in the stratum granulosum, accumulate keratohyalin and lipids before terminal differentiation. The \textbf{corneocytes} are the terminally differentiated keratinocytes, and are flattened, enucleated cells embedded in a lipid matrix, forming a robust barrier to mechanical stress and water loss. The \textbf{melanocytes},located in the basal layer,produce melanin to protect against ultraviolet radiation. And the \textbf{Langerhans cells}, dendritic immune cells within the suprabasal layers, act as antigen-presenting cells, linking cutaneous and systemic immunity.
The coordinated activity of these cell types enables the epidermis to function as a dynamic, multifunctional barrier that provides protection, pigmentation, sensory interface, and immunological defense.

\subsection{Dermis}
The dermis is the intermediate skin layer, composed mainly of fibrous connective tissue rich in collagen and elastin, which provide mechanical strength and elasticity~\cite{brown2022}. It can be divided into two main regions and performs key functions, including maintaining tensile integrity, supporting blood circulation, enabling oxygen and nutrient exchange, and contributing to immune defense~\cite{lee2024}. A dense capillary network within the dermis ensures continuous exchange of oxygen and metabolites between blood and tissue. The primary cellular component, the \textbf{fibroblasts}, synthesize collagen, elastin, and other extracellular matrix proteins, maintaining the dermis’s structural and mechanical properties. The endothelial cells line the capillaries, regulating the transport of gases and solutes, while \textbf{red blood cells} within these vessels are essential for oxygen delivery and thermo-regulation~\cite{brown2022}. From an electromagnetic perspective, fibroblasts and red blood cells are particularly relevant, as their dimensions and water content fall within the interaction regime of sub-THz and THz radiation. These features make them major contributors to scattering and absorption, thus influencing the overall dielectric behavior of the dermal tissue.

\subsection{Hypodermis}
The hypodermis is the deepest skin layer, mainly composed of adipose and connective tissues that provide structural support, cushioning, and energy storage~\cite{hirsch2023}. Its key functions include storing triglycerides for metabolic energy, minimizing heat loss to regulate body temperature, and protecting underlying organs from mechanical stress.  Among the cellular components of this layer, the \textbf{adipocytes} are the predominant cell type, storing energy in the form of triglycerides and contributing to thermal insulation and mechanical protection. A dense network of blood vessels runs between the dermis and hypodermis, facilitating nutrient and oxygen delivery. Within these vessels, \textbf{red blood cells} play a critical role in oxygen transport, supporting tissue metabolism and homeostasis. Overall, the hypodermis functions as a flexible, energy-rich, and protective layer that supports the upper skin strata while contributing to thermoregulation, mechanical stability, and vascular integration.

\subsection{Extracellular Matrix}

The extracellular matrix (ECM) is a complex network of macromolecules that provides structural and biochemical support to cells, maintaining tissue integrity, strength, and intercellular signaling. In the skin, it plays a key role in preserving architecture, elasticity, and hydration through water-binding components that regulate viscoelasticity and permeability~\cite{bosman2003functional,karamanos2021guide,watt2011cell}. ECM composition and organization vary across skin layers, reflecting their distinct structural and physiological functions. In the \textit{epidermis}, the ECM is confined mainly to the basement membrane, composed of collagen type IV, laminin, and proteoglycans that support cell adhesion and hydration. The \textit{dermis} is ECM-rich,  dominated by collagen I and III and elastin within a hydrated matrix of proteoglycans and glycosaminoglycans that retain bound water. The \textit{hypodermis}, in contrast, contains a looser ECM forming connective septa around adipose lobules, with collagen and elastin fibers interspersed with proteoglycans that provide mechanical support and local water retention around lipid-rich adipocytes. These structural variations govern the mechanical, hydraulic, and dielectric behavior of each skin layer, establishing the ECM as a key determinant of the tissue’s physical and electromagnetic properties across its full depth~\cite{krieg2011extracellular}.

\section{Computational Modeling of Cellular Permittivity in the sub-THz-to-THz Band}
\label{sec:cell_permittivity}
In this study, the complex permittivity of human cells and their surrounding ECM is evaluated over 100 GHz–1 THz, where dielectric behavior is dominated by water dynamics and protein/lipid relaxation. The model combines a multi-Debye relaxation framework with Maxwell–Garnett effective medium theory, treating cells as heterogeneous mixtures of water, proteins, and lipids within the ECM.

The multi-Debye model captures the multi-scale dynamics of cellular water and macromolecules in the Sub-THz/THz range. Since ions cannot follow the applied field above 100 GHz, ionic conductivity is neglected. The complex permittivity of each component (water, proteins, and lipids) is expressed as \cite{jornet}:
\begin{equation} \label{eq:debye_eq}
\varepsilon^*(\omega) = \varepsilon_{\infty} + 
\frac{\Delta\varepsilon_\alpha}{1 + j\omega\tau_{\alpha}} + 
\frac{\Delta\varepsilon_\beta}{1 + j\omega\tau_{\beta}} + + 
\frac{\Delta\varepsilon_\gamma}{1 + j\omega\tau_{\gamma}},
\end{equation}
where $\varepsilon_{\infty}$ is the high-frequency permittivity, and $\Delta\varepsilon_{\alpha,\beta,\gamma}$ and $\tau_{\alpha,\beta,\gamma}$ are the strengths and characteristic times of the three dominant relaxation processes~\cite{gabriel2018dielectric,wolf2012relaxation}. The \textit{$\alpha$-relaxation} arises from collective structural rearrangements in the hydration network~\cite{gabriel2018dielectric}, the \textit{$\beta$-relaxation} corresponds to more localized motions at higher frequencies~\cite{wolf2012relaxation,tannino2023heuristic}, and the \textit{$\gamma$-relaxation} represents ultrafast motions~\cite{wolf2012relaxation}. Proteins exhibit all three processes with a dominant $\alpha$ contribution, while lipids mainly show $\beta$- and $\gamma$-relaxations~\cite{jornet}.

Once the permittivity of each cellular component is defined, the effective permittivity of the whole cell, modeled as a water-based medium containing dispersed protein and lipid inclusions, is estimated using the Maxwell–Garnett effective medium theory~\cite{tannino2023heuristic}:
\begin{equation}
\tilde{\epsilon}(\omega) = \epsilon_\text{host}(\omega) + 3 \epsilon_\text{host}(\omega) \frac{\sum_{n=1}^N \phi_n \frac{\epsilon_n - \epsilon_\text{host}}{\epsilon_n + 2 \varepsilon_\text{host}}} {1 - \sum_{n=1}^N \phi_n \frac{\epsilon_n - \epsilon_\text{host}}{\epsilon_n + 2 \varepsilon_\text{host}}}, 
\label{eq:cell_permittivity_total}
\end{equation}
where $\epsilon_\text{host}$ and $\epsilon_n$ are the permittivity of the water (i.e. intracellular liquid) and of the cellular inclusions, respectively, and $\phi_n=\frac{m_{{n}}}{m_{\text{total}}}$ is the volumetric fraction of each inclusion based on its mass fraction $m_n$ with respect to the entire composite structure, $m_{\text{total}}$. Finally, the complex refractive index of the composite system is obtained as
\begin{equation}
n(\omega) = \sqrt{\tilde{\epsilon}(\omega)}=n'(\omega) - jn''(\omega).
\label{eq:refr_ind}
\end{equation}
It has a real part, governing the phase velocity, and an imaginary part, describing absorption~\cite{Calvo2021,Jepsen2019}. This approach provides the frequency-dependent optical properties for each cell type and the ECM~\cite{jornet,gabriel2018dielectric,wolf2012relaxation}.

\subsection{Cell Types and Properties}
We consider the main epidermal and dermal cell types, as well as blood and connective tissue cells. Each cell is modeled as a composite structure, with water as the host medium and proteins and lipids as dispersed inclusions. Table~\ref{tab:debye_parameters_reorg} lists the Debye relaxation parameters, while Table~\ref{tab:cell_properties} reports component concentrations in the considered skin cells~\cite{jornet,gabriel2018dielectric,wolf2012relaxation,tannino2023heuristic,madison2003barrier,van2014important}.
\begin{table}[!b]
\centering
\caption{Debye relaxation parameters of water, proteins, and lipids (100~GHz–1~THz, 37$^\circ$C)~\cite{jornet,gabriel2018dielectric,wolf2012relaxation,tannino2023heuristic}.}
\vspace{.2cm}
\label{tab:debye_parameters_reorg}
\begin{tabular}{@{}lccccccc@{}}
\toprule
\textbf{Component} & $\varepsilon_{\infty}$ & $\Delta\varepsilon_{\alpha}$ &$\Delta\varepsilon_{\beta}$ & $\Delta\varepsilon_{\gamma}$ &$\tau_{\alpha}$ (ps)  &$\tau_{\beta}$ (ps)  &$\tau_{\gamma}$ (ps) \\
\midrule
Water & 1.8 & 78 & -- & -- & 8.3 & -- & -- \\
Proteins & 2.0--2.5 & 10--50 &1--5 & 0.5--2.0 & 1--100 & 0.1--10 & 0.01--0.1 \\
Lipids & 2.0--2.2 & $\approx$~0 & 1--3 & 0.2--1.0 & -- & 0.1--1.0 & 0.01--0.05 \\
\bottomrule
\end{tabular}
\vspace{-0.5cm}
\end{table}
\begin{table}[!t] 
\centering 
\caption{Protein and lipid mass fractions and derived parameters for different skin cell types\cite{madison2003barrier}, \cite{van2014important}.} 
\label{tab:cell_properties}
\vspace{.2cm}
\begin{tabular}{|l|c|c|c|c|c|} 
\hline 
\textbf{Cell Type} & \textbf{Water (\%)} & \textbf{Protein (\%)} & \textbf{Lipid (\%)} \\ 
\hline 
Corneocytes & 10-15 & 70-80 & 10-15 \\ 
Granular Keratinocytes & 70-75 & 20-25 & 3-5\\ 
Spinous Keratinocytes & 75-80 & 15-20 & 2-3\\ 
Basal Keratinocytes & 75-80 & 15-20 & 2-3 \\ 
Langerhans Cells & 70-75 & 20-25 & 3-5 \\ 
Melanocytes & 75-80 & 15-20 & 2-3 \\ 
Red Blood Cells & 65-70 & 28-32 & 1-2 \\ 
Fibroblasts & 75-80 & 15-20 & 2-3 \\ 
Adipocytes & 10-20 & 5-10 & 70-85 \\ 
\hline
\end{tabular} 
\end{table}
Water exhibits a dominant $\alpha$-relaxation at ~8.3 ps, reflecting collective dipolar reorientation. Proteins display multiple relaxation modes with broader time distributions, including side-chain ($\beta$) and localized ($\gamma$) motions, while lipids contribute at shorter timescales due to restricted headgroup and tail mobility. Most skin components are smaller than 0.1~mm, making them much smaller than typical THz wavelengths. Therefore, scattering is negligible and the cell can be treated as an effective homogeneous dielectric inclusion. However, larger structures can induce THz wave scattering, which must be considered in both experimental measurements and modeling~\cite{Nikitkina2021}.

By combining Debye parameters with the volumetric fractions of cellular constituents in \eqref{eq:cell_permittivity_total}, the complex permittivity of each skin cell type is obtained and used to derive the corresponding frequency-dependent refractive index.

\subsection{Extracellular Matrix Composition and Dielectric Properties}
The ECM exhibits distinct biochemical and structural compositions across the epidermis, dermis, and hypodermis, which collectively determine the dielectric response of the tissue~\cite{krieg2011extracellular,watt2011cell}. In the \textit{epidermis}, water accounts for 65–70\% of the total mass, proteins (mainly collagen IV, laminin, and fibronectin) for 25–30\%, and lipids for 5\%, primarily in the stratum corneum. The \textit{dermis} is similarly hydrated, containing 65–75\% water, 20–30\% proteins (mainly collagen I/III and elastin), and a negligible lipid fraction (2\%). In the \textit{hypodermis}, the composition shifts toward adipose tissue with 20–30\% water, 5–10\% proteins, and 60–75\% lipids stored in adipocytes~\cite{karamanos2021guide}.

These compositional differences govern the dielectric behavior of each skin layer by influencing polarization and absorption, dominated by water in the epidermis and dermis, and by lipids in the hypodermis.  The effective permittivity of the ECM in each layer is computed using~\eqref{eq:cell_permittivity_total}, modeling it as a water-based medium with layer-dependent inclusion fractional volume, providing a comprehensive description of the skin’s electromagnetic behavior~\cite{tannino2023heuristic}.

\section{Intrabody Wave Propagation Losses}
\label{sec:intrabody_losses}
Understanding the dielectric properties of cells and the extracellular matrix is essential for accurate modeling of sub-THz and THz wave propagation in biological tissues. The total path loss within the human body arises from three frequency-dependent components: spreading, molecular absorption, and scattering losses. The overall attenuation factor is expressed as in~\cite{jornet}:
\begin{equation}
    L_{\text{tot}}(f) = L_{\text{spr}}(f) \times L_{\text{abs}}(f) \times L_{\text{sca}}(f),
\end{equation}
where $L_{\text{spr}}$ accounts for spherical wavefront spreading, $L_{\text{abs}}$ represents molecular absorption losses, and $L_{\text{sca}}$ denotes attenuation due to scattering processes.

\subsection{Spreading Loss}
Electromagnetic waves in biological tissues experience spreading losses, modeled as
\begin{equation}
L_{\text{spr}} = D \left(\frac{\lambda_g}{4\pi d}\right)^2 ,
\end{equation}
where $\lambda_g = \lambda/n'$ is the wavelength in the medium, $d$ the propagation distance, and $D$ the transmitting antenna directivity. The refractive index is defined as $n = n' - jn'' = \sqrt{\tilde{\varepsilon}}$, where $\tilde{\varepsilon}$ is the complex permittivity of both cells and ECM, obtained from~\eqref{eq:cell_permittivity_total}.

\subsection{Molecular Absorption Loss}
Molecular absorption arises from the interaction of THz waves with tissue constituents at their characteristic resonant frequencies~\cite{Saxena2024}. It is modeled using the Beer–Lambert law~\cite{Shi2022}:
\begin{equation}
L_{\text{abs}} = e^{-\mu_{\text{abs}} d},
\end{equation}
where $\mu_{\text{abs}}$ is the molecular absorption coefficient, given by $\mu_{\text{abs}} = 4\pi n'' / \lambda_g$~\cite{elayan2017multi}.

\subsection{Scattering Loss}
Light scattering in the skin originates from its microscopic heterogeneity and is evaluated \textit{cell by cell}. Rayleigh scattering applies to particles smaller than the wavelength, while Mie theory describes larger structures such as red blood cells or adipocytes~\cite{Cherkasova2021}. The scattering loss is expressed as
\begin{equation}
    L_{\text{sca}} = e^{-(\mu_{\text{sca}}^{\text{small}} + \mu_{\text{sca}}^{\text{large}}) d},
\end{equation}
where $\mu_{\text{sca}}^{\text{small}}$ and $\mu_{\text{sca}}^{\text{large}}$ are the scattering coefficients for small and large particles, respectively, accounting for tissue heterogeneity and avoiding the effective medium approximation~\cite{Cheong1990}.

For large particles, part of the incident energy is reflected while the rest is refracted. The total energy removed from the beam, the extinction has efficiency~\cite{vanDeHulst1981}
\begin{align}
Q_{\text{ext}} &= 2 - \frac{4}{p} \sin p + \frac{4}{p^2}(1 - \cos p), & 
Q_{\text{sca}}^{\text{large}} &= Q_{\text{ext}} - Q_{\text{abs}}.
\end{align}
where $p = 4\pi r (n - 1)/\lambda = 2(n - 1)\psi$ represents the phase delay through the particle center, and $\psi = 2\pi r / \lambda_g$ is the dimensionless size parameter.
The scattering coefficient for large particles is then computed as $ \mu_{\text{sca}}^{\text{large}} = \rho_{\text{large}}Q_{\text{sca}}^{\text{large}} \sigma_g,$
where  $\sigma_g = \pi r^2$ is the geometric cross section of a sphere  given the radius $r$ and $\rho_{\text{large}}$ is the particle number density, which can be written in terms of the volume fraction. 

For particles much smaller than the wavelength, the local electric field induces an oscillating dipole, leading to \textit{Rayleigh scattering}~\cite{Cherkasova2021}.
The scattering efficiency of small spherical absorbing particles is given by \cite{jornet}
\begin{equation}
Q_{\text{sca}}^{\text{small}} = \frac{8}{3} \psi^4 \text{Re} \left( \frac{n^2 - 1}{n^2 + 2} \right)^2,
\end{equation}
In this case, the scattering coefficient for small particles is $ \mu_{\text{sca}}^{\text{small}} = \rho_{\text{small}} Q_{\text{sca}}^{\text{small}} \sigma_g$.

Overall, attenuation in biological tissues is highly frequency-dependent, arising from the combined effects of spreading, molecular absorption, and scattering across multiple spatial scales.

\section{Simulation framework}
\label{sec:simulation_framework}
Human skin tissue is simulated using a MATLAB-based voxelized framework. The tissue is discretized with voxels of size $dx = 0.01~\mathrm{mm}$ (10~µm) over a lateral extent of $X = Y = 0.1~\mathrm{mm}$ and depth $Z = 5.0~\mathrm{mm}$. Cells are modeled as spheres with layer-specific diameters and concentrations. Table~\ref{tab:cell_all} summarizes the cell concentrations $\rho$, diameters $d$, and an average z-intervals $Z_\mathrm{layer}$ corresponding to their natural layer positions. The number of cells per layer is limited by literature-reported concentrations, while ECM concentration is determined from the remaining volume not occupied by cells.
\begin{table}[!b]
\centering
\caption{Cell Parameters in the Skin Model \cite{nikitkina2021terahertz},\cite{graham2019human}}\label{tab:cell_all}
\vspace{.2cm}
\begin{tabular}{l c c c}
\hline
Cell Type &  $\rho$ [cells/mm$^3$] &  $d$ [µm] &  $Z_\mathrm{layer}$ [µm] \\
\hline
Corneocytes & $1.0 \times 10^7$ & 23 & 0--20 \\
Granular Keratinocytes & $0.8 \times 10^6$ & 15 &20--50 \\
Spinous Keratinocytes & $0.7 \times 10^6$ & 15 &50--120 \\
Melanocytes & $2.0 \times 10^4$ & 14 &50--120 \\
Basal Cells & $1.0 \times 10^6$ & 8 &120--130 \\
Langerhans Cells & $3.0 \times 10^4$ & 10 &50--120 \\
Merkel Cells & $1.0 \times 10^3$ & 10 &50--120 \\
Fibroblasts & $5.0 \times 10^4$ & 20 &130--3000 (Dermis) \\
Red Blood Cells & $6.0 \times 10^5$ & 8 & Capillaries / Arterioles \\
Adipocytes & $1.0 \times 10^5$ & 80 & 3000+ (Hypodermis) \\
\hline
\end{tabular}
\end{table}
\subsection{Statistical Skin Scenario Definition}
\begin{figure}[!b]
    \centering
    \begin{subfigure}[b]{0.32\textwidth}
        \centering
        \includegraphics[width=0.8\textwidth]{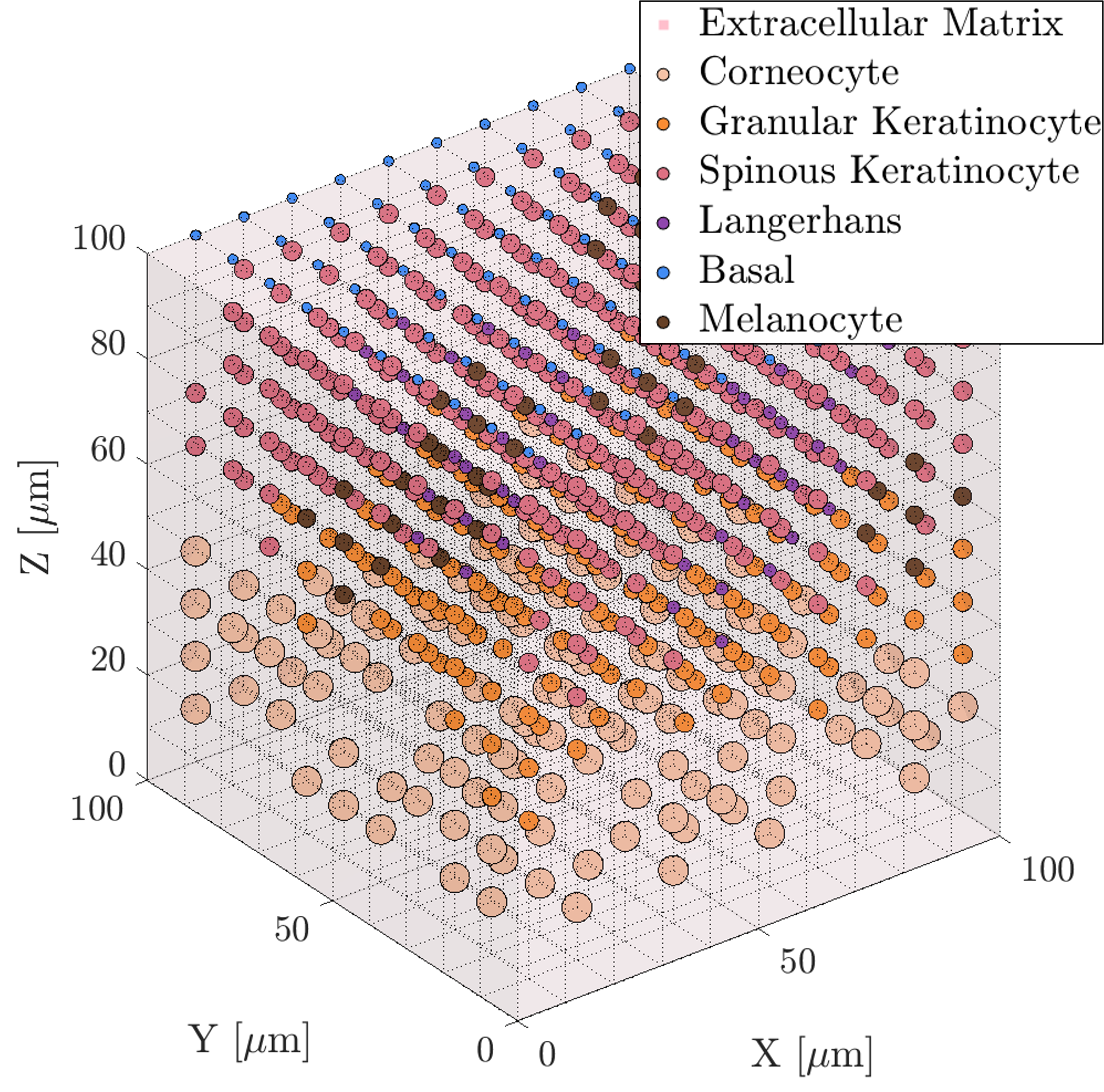}
        \caption{}
        \label{fig:subfig1}
    \end{subfigure}
    % --- Second subfigure ---
    \begin{subfigure}[b]{0.32\textwidth}
        \centering
        \includegraphics[width=0.8\textwidth]{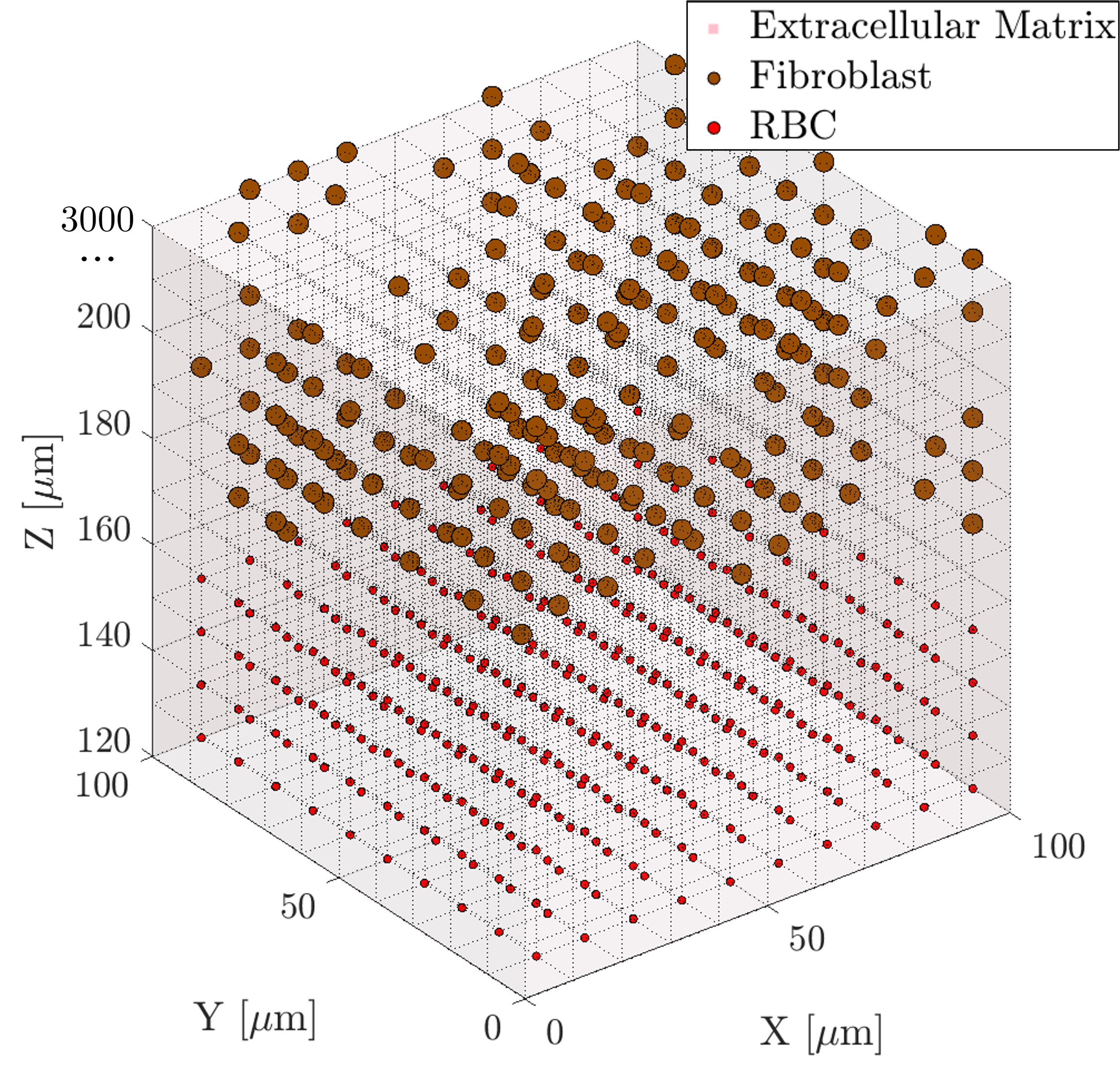}
        \caption{}
        \label{fig:subfig2}
    \end{subfigure}
    \hfill
    % --- Third subfigure ---
    \begin{subfigure}[b]{0.32\textwidth}
        \centering
        \includegraphics[width=0.8\textwidth]{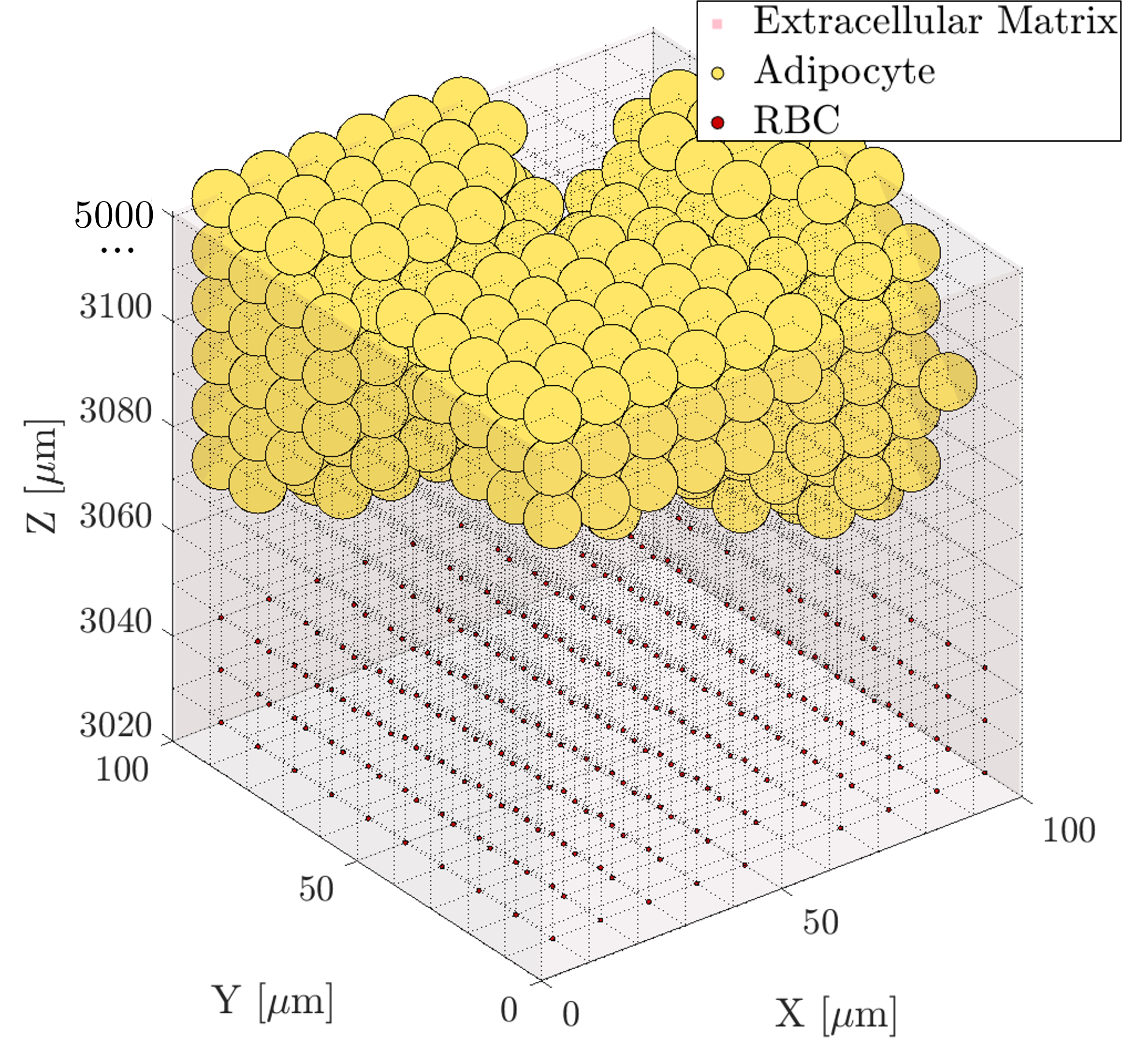}
     \caption{}
    \label{fig:subfig3}
    \end{subfigure}
    \caption{Voxel-based 3D representation of skin with randomly distributed spherical cells, scaled by diameter, in (a) epidermis, (b) dermis, and (c) hypodermis.}
\end{figure}
In the simulation, cells are modeled as spheres within their respective tissue layer volumes ($V_{\text{layer}}$). Each cell’s position, $\mathbf{p} = [x,y,z]^{\mathrm{T}}$, is assigned via a \emph{disc Poisson process}, which ensures a minimum center-to-center distance equal to the sum of the radii of any two cells:
\begin{equation}
\mathcal{P} = \Big\{ \mathbf{p}_i \in V_\mathrm{layer} \,\big|\, \|\mathbf{p}_i - \mathbf{p}_j\| \ge \frac{d_i}{2} + \frac{d_j}{2},~\forall j \neq i \Big\}.
\end{equation}
This defines a spatial probability distribution for the center of each cell in $V_\mathrm{layer}$. 

Vessels are represented as line segments inside the $V_\mathrm{layer}$ and with centers and orientations sampled from such a \emph{line Poisson process}. It distributes lines or line segments in space according to a homogeneous Poisson point process over a parameter space defining position and orientation~\cite{kingman1992poisson}. Lateral coordinates $(x_v, y_v)$ of each vessel center are drawn from a homogeneous 2D Poisson point process over the layer plane, while the vertical coordinate $z_v$ spans the layer thickness. Each segment is aligned along the vessel axis, with length determined by vessel type. Formally, the set of vessel center positions is
\begin{equation}
\mathcal{L} = \Big\{ \mathbf{p}_v = (x_v, y_v, z_v) \in V_{\mathrm{layer}},\; v = 1, \dots, N_{\mathrm{vessels}} \Big\},
\end{equation}
where $N_\mathrm{vessels}$ is the total number of vessels of a given type. This approach produces a spatially homogeneous, statistically independent distribution of vessel centers in the lateral plane, while allowing control over vessel density, orientation, and length. Anatomical constraints are incorporated: dermal capillaries are confined to the upper layer near the epidermal–dermal interface, reflecting the dense superficial microvasculature, whereas vessels in the hypodermis span the full layer depth, consistent with the more uniform distribution of larger arterioles and venules.

Red blood cells are then randomly distributed within the vessels using a \emph{disc Poisson process}, consistent with the vessel geometry and avoiding overlap.

Figure~\ref{fig:subfig1} shows a representative realization of cell placement in the epidermis, while Figures~\ref{fig:subfig2} and~\ref{fig:subfig3} depict the corresponding distributions in the dermis and hypodermis, respectively.

\section{Numerical Results}
\begin{figure}[!b]
    \centering
        \begin{subfigure}{0.48\linewidth}\label{subfig:attenuation100}
        \centering
        \includegraphics[width=\linewidth]{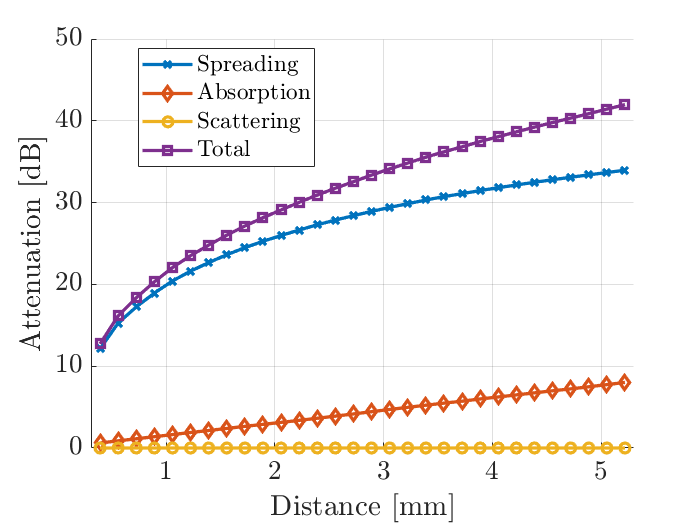}
        \caption{}
    \end{subfigure}
    \hfill
    \begin{subfigure}{0.48\linewidth}\label{subfig:attenuation1000}
    \centering
    \includegraphics[width=\linewidth]{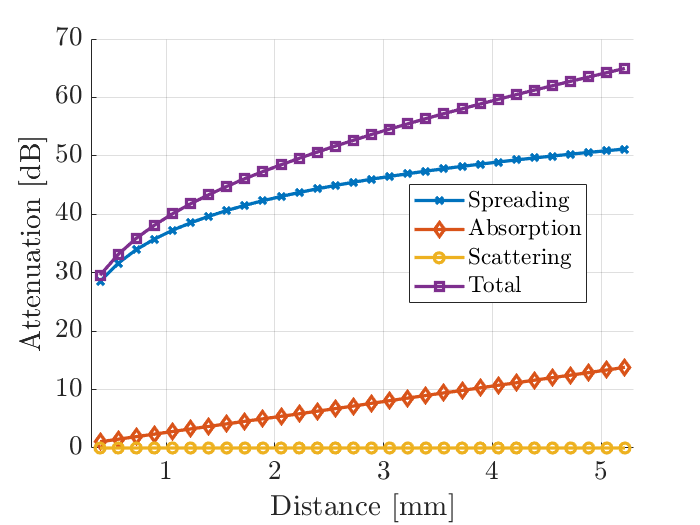}
    \caption{}
    \end{subfigure}
    \caption{Attenuation due to spreading, absorption, scattering, and the total vs distance, for (a) $f = 100$ GHz and (b) $f = 1$ THz.}
    \label{attenuation}
\end{figure}

This section quantitatively analyzes THz propagation in multilayer skin models, highlighting frequency-dependent attenuation, absorption, and scattering at 100~GHz and 1~THz.
Figures~\ref{attenuation}a–b show the relative contributions of attenuation mechanisms from Section~\ref{sec:intrabody_losses} at 100~GHz and 1~THz. The skin’s heterogeneous structure produces frequency-dependent losses. Spreading loss ($L_{\mathrm{spr}}$) dominates, reflecting its reliance on the effective wavelength, contributing 35~dB over 5~mm at 100~GHz and 40~dB at 1~THz. Absorption loss ($L_{\mathrm{abs}}$) is smaller at 100~GHz (8~dB) but rises to 15~dB at 1~THz due to stronger molecular absorption by water and biomolecules. These results highlight the importance of frequency selection in sub-THz and THz biomedical applications: higher frequencies enhance interactions with tissue water and molecular resonances, markedly affecting propagation.
Scattering ($L_{\mathrm{sca}}$) remains minor at both frequencies but shows clear frequency dependence. At 100~GHz, wavelengths are much larger than cells, producing low, uniform Rayleigh scattering. At 1~THz, wavelengths approach cellular scales, shifting toward Mie scattering with stronger layer-specific effects. Total path loss sums spreading, absorption, and scattering, reaching 45~dB over 5~mm at 100~GHz and $\>$65~dB at 1~THz, reflecting greater sensitivity of higher frequencies to tissue losses.

\begin{figure}[!t]
    \centering
        \begin{subfigure}{0.48\linewidth}\label{subfig:abs100}
        \centering
        \includegraphics[width=\linewidth]{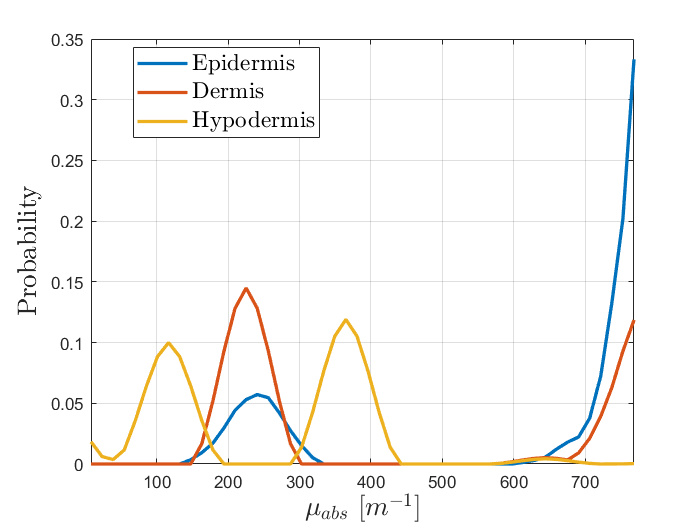}
        \caption{}
    \end{subfigure}
    \hfill
    \begin{subfigure}{0.48\linewidth}\label{subfig:abs1000}
    \centering
    \includegraphics[width=\linewidth]{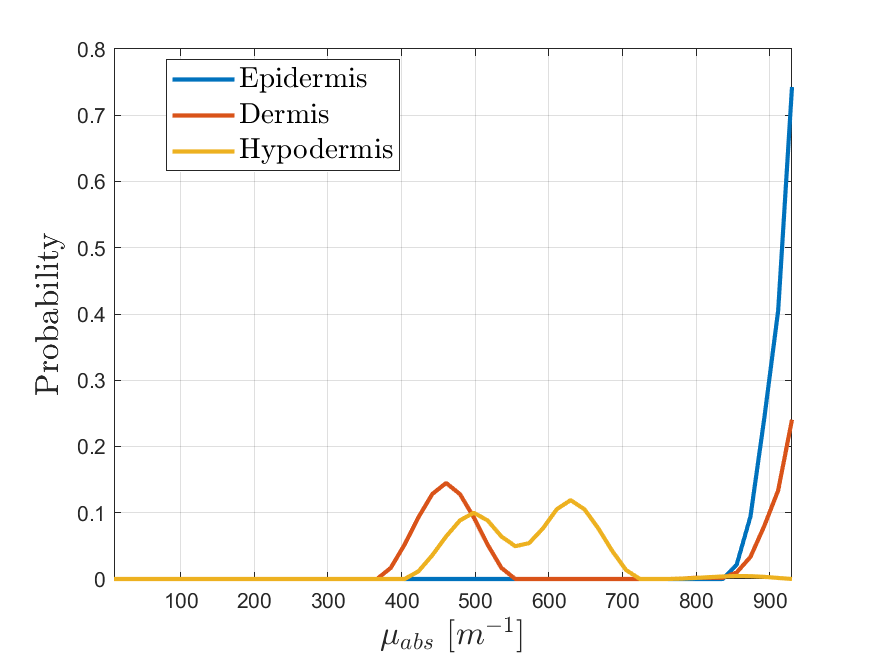}
    \caption{}
    \end{subfigure}
    \caption{Absorption coefficient probability density function for (a) $f = 100$ GHz and (b) $f = 1$ THz.}
    \label{absorption_MATLAB}
\end{figure}
\begin{figure}[!b]
    \centering
        \begin{subfigure}{0.48\linewidth}\label{subfig:sca100}
        \centering
        \includegraphics[width=\linewidth]{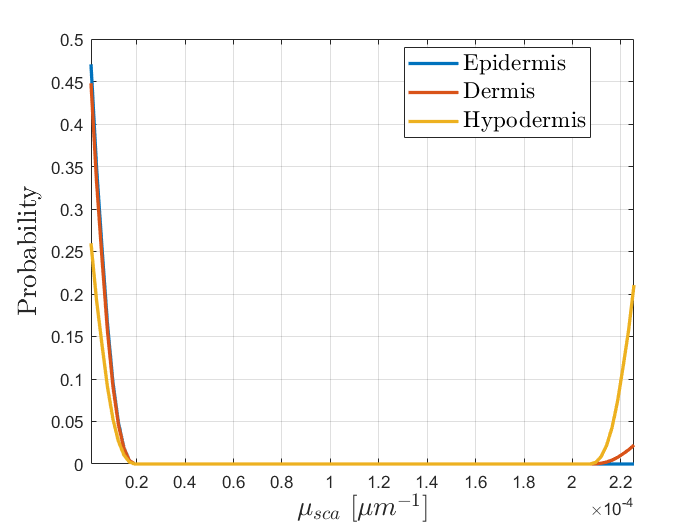}
        \caption{}
    \end{subfigure}
    \hfill
    \begin{subfigure}{0.48\linewidth}\label{subfig:sca1000}
    \centering
    \includegraphics[width=\linewidth]{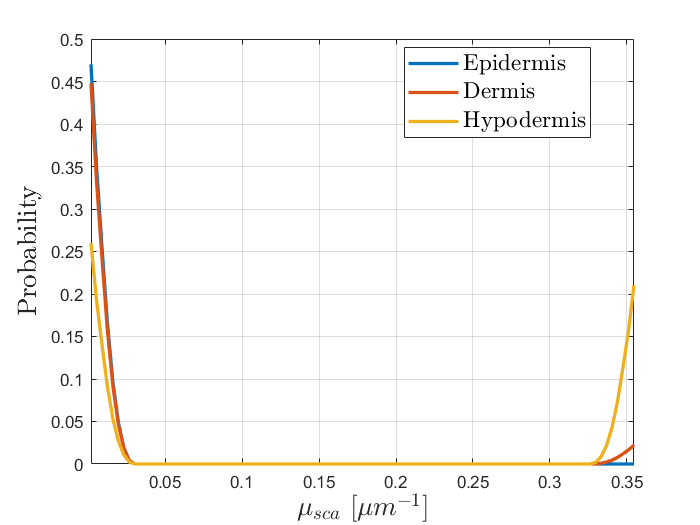}
    \caption{}
    \end{subfigure}
    \caption{Scattering coefficient for (a) $f = 100$ GHz and (b) $f = 1$ THz.}
    \label{scattering}
\end{figure}

Tissue-specific absorption is shown in Figs.~\ref{absorption_MATLAB}, with frequency-dependent trends reflecting each skin layer’s composition. At 100~GHz, the epidermis shows a high absorption peak near 700--800~m\(^{-1}\) reflecting its high water content and strong dielectric losses, the dermis peaks around 200--300~m\(^{-1}\) from its collagen--water mix, and the hypodermis exhibits a broader, bimodal distribution (\(\sim 100\) and 350~m\(^{-1}\)) reflecting its adipose/connective tissue heterogeneity. At 1~THz, distributions shift higher, indicating increased losses due to enhanced dipolar relaxation of water molecules and stronger frequency-dependent coupling. The epidermis peaks sharply near 900~m\(^{-1}\), the dermis centers around 500~m\(^{-1}\), and the hypodermis ranges from 400--700~m\(^{-1}\), showing moderate absorption. These results highlight how skin absorption depends on frequency and tissue composition: water-rich layers dominate, while lipid-rich hypodermis shows weaker, more variable absorption. Such probabilistic characterization is crucial for realistic THz propagation models accounting for tissue heterogeneity.

Tissue-specific scattering properties are shown in Fig.~\ref{scattering}, where the probability distributions of the scattering coefficient (\(\mu_{\mathrm{sca}}\)) reflect structural rather than compositional differences of each layer. At 100~GHz, all layers exhibit extremely low scattering, with sharp peaks near $2 \cdot 10^{-5}~\mu\mathrm{m}^{-1}$. The epidermis and dermis show nearly identical distributions with probability densities above 0.45, indicating minimal Rayleigh scattering from cells and collagen fibers (\(\lambda \approx 3~\mathrm{mm} \gg 1\text{--}100~\mu\mathrm{m}\) scatterers dimension). The hypodermis shows a similar trend with a slightly broader tail up to $2 \cdot 10^{-4}~\mu\mathrm{m}^{-1}$ due to larger adipocytes. At 1~THz, distributions broaden as Mie scattering emerges (\(\lambda \approx 300~\mu\mathrm{m}\)), with epidermis and dermis developing extended tails up to 0.25--0.35~\(\mu\mathrm{m}^{-1}\) at low probabilities (\(\sim 0.01\text{--}0.02\)). The hypodermis shows a pronounced bimodal distribution with peaks near 0 and 0.01--0.35~\(\mu\mathrm{m}^{-1}\), reflecting its heterogeneous adipocyte content. Overall, scattering remains a secondary loss mechanism compared to absorption, even at THz frequencies.

These results show that scattering becomes more relevant at higher frequencies, where cellular and subcellular heterogeneities interact more effectively with the electromagnetic field.  However, the persistent dominance of low-scattering probabilities across all layers indicates that spreading, followed by absorption remain the primary attenuation mechanisms in skin tissue throughout the subTHz and THz regimes.

This numerical analysis provides a quantitative understanding of THz interaction with multilayer skin. Lower frequencies (100~GHz) exhibit lower attenuation and deeper penetration, suitable for probing deeper tissues. Instead, higher frequencies (1~THz) increase absorption but offer better spatial resolution and contrast, enhancing sensitivity to superficial and structural variations. The complementary properties of these regimes suggest that hybrid or dual-frequency approaches, combined with adaptive power management, could improve THz imaging and diagnostics.

\section{Conclusion}
This work presents a comprehensive dielectric modeling framework for human skin in the sub-THz and THz ranges. By combining multi-Debye relaxation with effective medium theory, the model captures frequency-dependent permittivity of cellular constituents and the extracellular matrix across the epidermis, dermis, and hypodermis. Physiologically realistic parameters, including water content, protein and lipid fractions, and cellular heterogeneity, enable a physically interpretable description of tissue dielectric properties. Numerical simulations using a voxel-based statistical skin model reveal the distinct contributions of spreading, molecular absorption, and scattering to wave attenuation. At lower sub-THz frequencies, spreading dominates, allowing deeper penetration, while at higher THz frequencies, absorption and scattering increase, improving tissue contrast and layer differentiation.

Overall, this study provides a physically grounded framework for predicting electromagnetic interactions in skin, offering practical insights for designing and optimizing next-generation sub-THz and THz imaging and sensing systems.

\bibliography{ref}
\bibliographystyle{IEEEtran}

\end{document}